\documentclass[11pt,twoside]{article}

\usepackage{asp2006}

\begin{document}
\title{High-Redshift Quasars at the Highest Resolution: VSOP Results}
\author{S. Frey,$^{1,2}$ L.I. Gurvits,$^{3}$, A.P. Lobanov,$^{4}$ R.T. Schilizzi,$^{5}$ and Z. Paragi$^{3,2}$}
\affil{
$^1$ F\"OMI Satellite Geodetic Observatory, P.O. Box 585, H-1592 Budapest, Hungary\\
$^2$ MTA Research Group for Physical Geodesy and Geodynamics, P.O. Box 91, H-1521 Budapest, Hungary\\
$^3$ Joint Institute for VLBI in Europe, Postbus 2, 7990 AA Dwingeloo, The Netherlands\\
$^4$ Max-Planck-Institut f\"ur Radioastronomie, Auf dem H\"ugel 69, D-53121 Bonn, Germany\\
$^5$ International Square Kilometre Array Project Office, Postbus 2, 7990 AA Dwingeloo, The Netherlands\\
}

\begin{abstract}
We studied the radio structure of high-redshift ($z>3$) quasars with VSOP at 1.6 and 5~GHz. These sources are the most distant objects ever observed with Space VLBI, at rest-frame frequencies up to $\sim25$~GHz. Here we give an account of the observations and briefly highlight the most interesting cases and results. These observations allowed us, among other things, to estimate the mass of the central black holes powering these quasars, to identify large misalignments between the milli-arcsecond (mas) and sub-mas scale radio structures, and to detect apparent superluminal motion at sub-mas scale.
\end{abstract}

\section{Introduction}

Twenty of the most distant ($z>3$) radio quasars have been proposed for Space Very Long Baseline Interferometry (SVLBI) observations at either 1.6 or 5~GHz or both in the VLBI Space Observatory Programme \citep[VSOP;][]{hira00}. These were the brightest known, with total flux densities higher than $\sim400$~mJy. Our goal was to study the structural properties of high-redshift sources at the highest possible resolution at emitted frequencies up to $\sim20-25$~GHz. This allowed us to investigate source compactness, relativistic jet propagation, and eventually to estimate physical parameters of the jets and the active nuclei feeding them.

\section{VSOP observations}
More than a half of the 44 different experiments proposed could be scheduled and successfully observed between 1997 and 2002. The quality of the data from individual experiments varied, depending on the length of the observations and the ($u,v$) coverage which was determined by the co-observing ground radio telescope network and the inclination of the satellite orbital plane with respect to the radio source direction. 

All the sources successfully targeted were detected and resolved at the longest VSOP baselines, with typical correlated flux densities less than $\sim200$~mJy \citep{gurv00}. The majority of the sources showed compact core-dominated radio structures. Our results for individual high-redshift sources have been published in a series of papers over the past decade \citep{hira98,gur+00,gurv00,frey00,fre+00,loba01,frey02,gurv03,yang06}.

The highest-redshift object observed with VSOP in this project was the quasar 1508+572 ($z=4.30$). Given the observing frequency of 5~GHz, this was already almost a ``millimetre SVLBI'' experiment, as far as the wavelength of the detected radiation in the rest frame of the source ($\lambda \approx 1$~cm) is concerned. We briefly review some of our major findings below.

\section{Project highlights}

The X-ray absorption of the quasar 0014+813 \citep[$z=3.37$;][]{hira98,gur+00} is known to be higher than the Galactic value. The excess absorption is thought to be intrinsic to the quasar. This suggests that we see a jet which lies close to the plane of the sky. On the other hand, the prominent core--jet structure visible in our SVLBI image indicates highly beamed radio emission originating from a jet pointing nearly to the line of sight. This apparent contradiction could be resolved if the compact radio structure of this quasar is sharply bent at sub-mas angular scales. Other sources with misaligned radio jets between mas and 10-mas or larger scales were also found in our sample, e.g. 0201+113 \citep[$z=3.61$;][]{fre+00}, 1351$-$018 \citep[$z=3.71$;][]{frey02} and 1354$-$174 \citep[$z=3.15$;][]{gur+00,frey00}. 

Constrained by the lifetime of the satellite, there was only a limited possibility for longer-term monitoring observations. However, in the case of 1351$-$018, we could obtain 5-GHz VSOP data in April 1998 and January 2001 as well. The extremely high angular resolution of the SVLBI network allowed us to identify a jet component within $\sim1$~mas from the core and to measure its apparent outward proper motion as $\mu = 0.18 \pm 0.07$~mas/yr. Based on the measurements at just two epochs separated by less than 3 years in time, this could be considered a tentative detection of apparent superluminal motion of $\sim10 c$ at an extremely high redshift.

With the help of the long SVLBI baselines, it may become possible to resolve the transverse structure of the parsec-scale jets. Since the jet spectra are steep, it is particularly important for such a study that a relatively low observing frequency (1.6~GHz) was available in VSOP, allowing us to see prominent jet structures in some sources even at these high redshifts. The lower angular resolution due to the long wavelength is compensated by the long ground--space VLBI baselines. A stunning example is the quasar 2215+020 \citep[$z=3.57$;][]{loba01} with an exceptionally long radio jet at this high redshift that could be traced up to 80~mas at 1.6~GHz. Under plausible assumptions on the values of magnetic field inside and outside the jet, and using the jet cross-section size directly measured by SVLBI, the mass of the central black hole driving the activity of the quasar can be estimated. For 2215+020, we obtained the value of $4 \times 10^9$~$M_{\odot}$. Similar considerations could be applied for the quasars 1442+101 \citep[$z=3.52$;][]{gurv03} and 1402+044 \citep[$z=3.19$;][and theese proceedings]{yang06}, leading to rough mass estimates in the order of $\sim 10^8-10^9$~$M_{\odot}$.

\acknowledgements
We gratefuly acknowledge the VSOP Project, which was led by the Institute of Space and Astronautical Science (Japan) in cooperation with many organizations and radio telescopes around the world. We thank the participation of K.\'E. Gab\'anyi, H. Hirabayashi, X. Hong, N. Kawaguchi, K.I. Kellermann, H. Kobayashi, Y. Murata, I.I.K. Pauliny-Toth and J. Yang in various stages of this project.
SF acknowledges partial support received from the Hungarian Scientific Research Fund (OTKA T046097) and the Hungarian Space Office (TP-314).


\begin{thebibliography}{}

\bibitem[Frey(2000)]{frey00}
Frey, S. 2000, VLBI Studies of Extremely Distant Quasars, PhD Thesis (Budapest: E\"otv\"os University) 

\bibitem[Frey et al.(2000)]{fre+00}
Frey, S., Gurvits, L.I., Schilizzi, R.T., et al. 2000, in Proc. 5th EVN Symposium, ed.
J.E. Conway, A.G. Polatidis, R.S. Booth \& Y.M. Pihlstr\"om (G\"oteborg: Chalmers Technical University), 41 

\bibitem[Frey et al.(2002)]{frey02}
Frey, S., Gurvits, L.I., Lobanov, A.P., et al. 2002, in Proc. 6th EVN Symposium, ed.
E. Ros, R.W. Porcas, A.P. Lobanov \& J.A. Zensus (Bonn: MPIfR), 89 

\bibitem[Gurvits(2000)]{gurv00}
Gurvits, L.I. 2000, in Astrophysical Phenomena Revealed by Space VLBI, ed. 
H. Hirabayashi, P.G. Edwards \& D.W. Murphy (Sagamihara: ISAS), 151 

\bibitem[Gurvits(2003)]{gurv03}
Gurvits, L.I. 2003, in Active Galactic Nuclei: from Central Engine to Host Galaxy, ASP Conference Series, Vol. 290, ed. 
S. Collin, F. Combes \& I. Shlosman (San Francisco: ASP), 51

\bibitem[Gurvits et al.(2000)]{gur+00}
Gurvits, L.I., Frey, S., Schilizzi, R.T., et al. 2000, Adv. Space Res., 26, 719

\bibitem[Hirabayashi et al.(1998)]{hira98}
Hirabayashi, H., Hirosawa, H., Kobayashi, H., et al. 1998, \science, 281, 1825 

\bibitem[Hirabayashi et al.(2000)]{hira00}
Hirabayashi, H., Hirosawa, H., Kobayashi, H., et al. 2000, \pasj, 52, 955 

\bibitem[Lobanov et al.(2001)]{loba01}
Lobanov, A.P., Gurvits, L.I., Frey S., et al. 2001, \apj, 547, 714

\bibitem[Yang et al.(2006)]{yang06}
Yang, J., Gurvits, L.I., Lobanov, A.P., Frey, S., \& Hong, X. 2006, Proceedings of Science, PoS(8thEVN)086

\end{thebibliography}
\end{document}